\documentclass[twocolumn]{aastex631}

\shorttitle{ }
\shortauthors{Kalari et al.}

\begin{document}

\title[Resolving R136 in the optical]{Resolving the core of R136 in the optical}

\author{Venu M. Kalari}
\affiliation{Gemini Observatory/NSF’s NOIRLab, Casilla 603, La Serena, Chile}
\affiliation{Departamento de Astronomia, Universidad de Chile, Casilla 36-D, Santiago, Chile}

\author{Elliott P. Horch}
\affiliation{Department of Physics, Southern Connecticut State University, 501 Crescent Street, New Haven, CT 06515, USA}

\author{Ricardo Salinas}
\affiliation{Gemini Observatory/NSF’s NOIRLab, Casilla 603, La Serena, Chile}

\author{Jorick S. Vink}
\affiliation{Armagh Observatory and Planetarium, College Hill, Armagh, BT61 9DG, UK}

\author{Morten Andersen}
\affiliation{Gemini Observatory/NSF’s NOIRLab, Casilla 603, La Serena, Chile}
\affiliation{European Southern Observatory, Karl-Schwarzschild-Strasse 2, 85748 Garching bei M{\"u}nchen, Germany}

\author{Joachim M. Bestenlehner}
\affiliation{Department of Physics Astronomy, University of Sheffield, Hounsfield Road, Sheffield, S3\,7RH, UK}

\author{Monica Rubio}
\affiliation{Departamento de Astronomia, Universidad de Chile, Casilla 36-D, Santiago, Chile}



\begin{abstract}

The sharpest optical images of the R136 cluster in the Large Magellanic Cloud are presented, allowing for the first time to resolve members of the central core, including R136a1, the most massive star known. These data were taken using the {\it Gemini} speckle imager Zorro in medium-band filters with effective wavelengths similar to {\it BVRI} achieving angular resolutions between 30--40\,{\it mas}. All stars previously known in the literature, having $V<$16\,mag within the central 2$\arcsec\times2\arcsec$ were recovered. Visual companions ($\geq$40\,mas; 2000 au) were detected for the WN5h stars R136\,a1, and a3. Photometry of the visual companion of a1 suggests it is of mid O spectral type. Based on new photometric luminosities using the resolved Zorro imaging, the masses of the individual WN5h stars are estimated to be between 150-200\,$M_{\odot}$, lowering significantly the present-day masses of some of the most massive stars known. These mass estimates are critical anchor points for establishing the stellar upper-mass function.

\end{abstract}



\section{Introduction} \label{sec:intro}

\begin{figure*}
\plotone{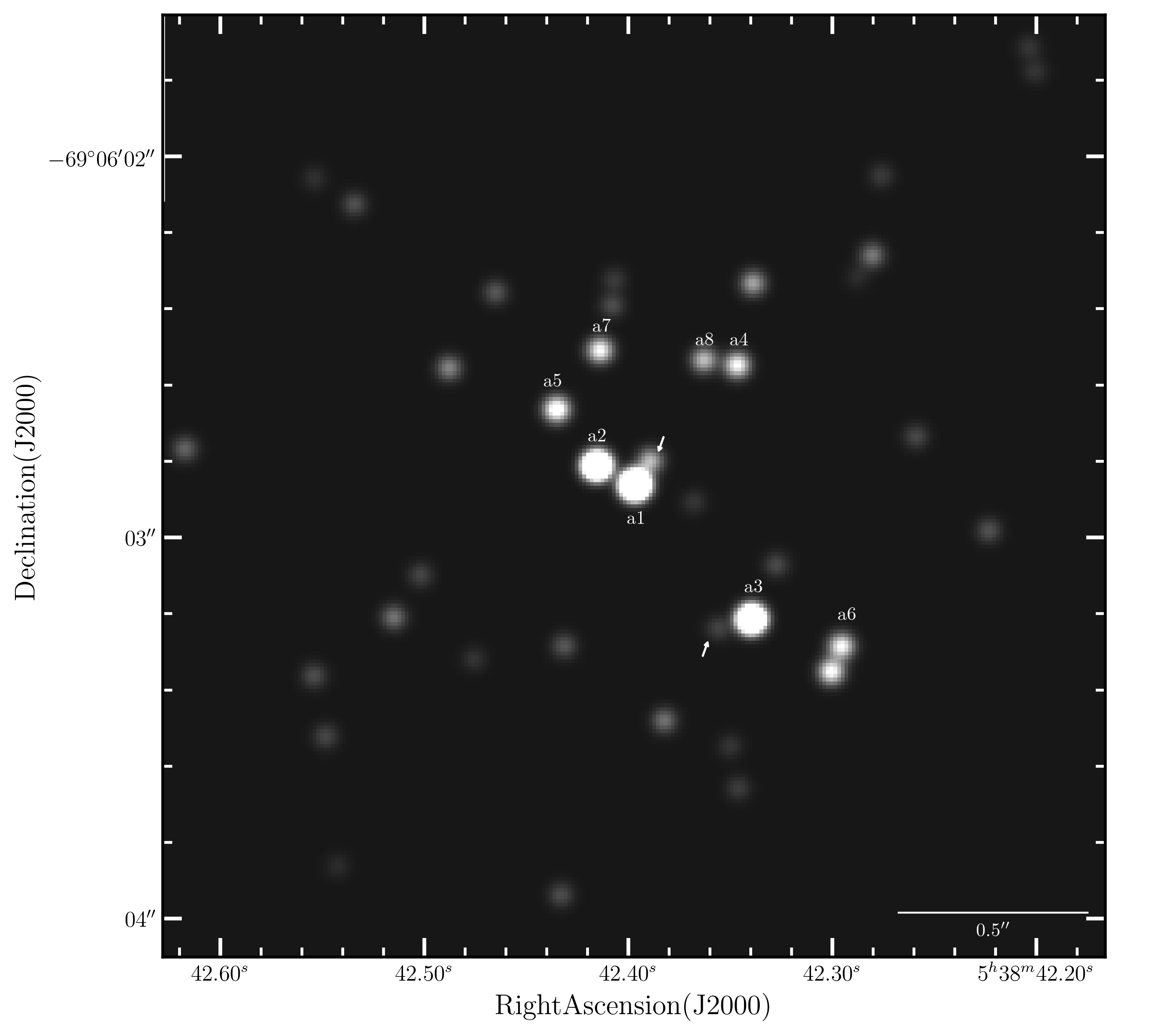}
\caption{Zorro EO832 imaging of R136. R136 resolved stars from \cite{Baier} are marked. Arrows mark the resolved companions to the WN5h stars R136\,a1, and a3. North is up and east is to the left}  
\label{zorro}
\end{figure*}

\begin{figure*}
\plotone{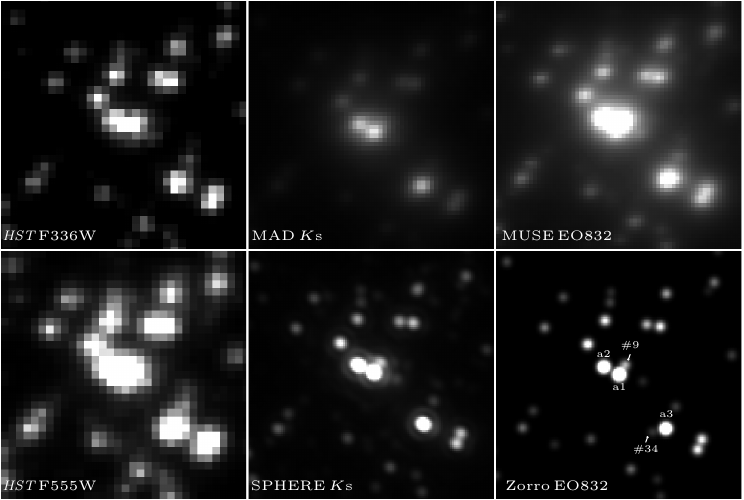}
\caption{Grayscale thumbnails of the central core of R136 (0.8$\arcsec\times0.8\arcsec$) centerd on R136a1, at highest available angular resolutions in the optical and infrared (IR). The instrument or telescope, and the filter of the images are given in the lower left hand corner of each thumbnail, and from top to bottom, left to right are {\it HST} F336W, and F555W images \citep{Hunter95}, IR adaptive optics (AO) images from the MAD and SPHERE instrument taken at the VLT in $K$s (\citealt{Camp10};\citealt{Khor21}), optical AO imaging using the integral field spectrograph MUSE \citep{muse}, and the Zorro images. Apparent is the improved angular resolution from AO imaging in IR, and the vastly improved resolution from Zorro compared to other optical images. All images are in linear scale, with intensity chosen to best separate the central components visually. }  
\label{comp}
\end{figure*}

The formation of very massive stars ($M_{\ast}>100\,M_{\odot}$) is an unsolved problem in astrophysics. Constraining the upper-mass limit, and binarity is crucial to our understanding of their formation. Most massive stars reside within the dense central cores of extincted young massive star clusters, and have short lifetimes ($\lesssim$2–3 Myr), making observations challenging. Particularly challenging are resolving local templates of rich super-star clusters, with cluster masses more than $10^4\,M_\odot$, that were common in the early Universe. These are highly compact, relatively distant objects making extreme spatial resolution observations crucial to resolve individual very massive stars found at their centers. 

The Large Magellanic Cloud (metallicity, $Z\sim1/2Z_{\odot}$) cluster R136 offers the best local template for starburst clusters found at higher redshifts, and is the most massive resolved star cluster known. It has a stellar mass of $\sim10^4\,M_{\odot}$ within its central 0.2\,pc core \citep{Crow16}, containing a total $\sim$25 O type stars within 1$\arcsec$ ($\sim$0.2 parsecs at a distance of 49.6\,kpc; \citealt{distance}). The most massive star yet known-- R136a1 is situated at it’s center, and has an estimated initial mass between 250–320\,$M_{\odot}$ \citep{Best20}, setting the presently known empirical stellar upper-mass limit (cf. \citealt{figer, kroupa}).

The observational challenges to resolve the central core of R136 and estimate the mass of individual stars, even with current instrumentation is significant. Speckle interferometry by \cite{Baier} first resolved the central core of R136 into eight components (marked in Figure\,\ref{zorro}). {\it Hubble Space Telescope} ({\it HST}) optical (\citealt{Hunter95}) and infrared (\citealt{Camp10,Khor21}) imaging confirmed this (Figure\,\ref{comp}). {\it HST} ultraviolet (UV) spectroscopy was later used by \cite{Crow16} to determine the effective temperature of R136a1 (WN5h spectral type), and an initial mass estimate of 325\,$M_{\odot}$. A more recent study by \cite{Best20} used updated stellar atmosphere models, which in particular include more detailed physics of the N{\scriptsize V} lines, to better to estimate a lower effective temperature (although some difficulties persisted with fitting the N{\scriptsize V} lines). In conjunction, they used observations \citep{Khor21} from the SPHERE instrument at the Very Large Telescope (VLT), which offers better extinction correction especially given the cloud content of this region \citep{Kalari18}. These yielded high-resolution adaptive optics (AO) near infrared (IR) Ks magnitudes that led to a mass of 251\,$M_{\odot}$. The angular resolution of {\it HST} ($\sim$50--60\,{\it mas}) does not resolve the central core of the cluster completely (Figure\,\ref{comp}). Unresolved photometry may lead to discrepancy in mass estimates as it gives rise to varying luminosity estimates, even under similar assumptions, with log\,$L/L_{\odot}$ varying between 6.33--6.94\,dex from {\it HST} and AO IR photometry. Comparing optical and IR thumbnails of the core of R136, centered on R136a1 in Figure\,\ref{comp}, the need for higher angular resolution to resolve the central core is apparent. 

In this paper, we present deep speckle images of R136 cluster in medium-band filters having central wavelengths at 466, 562, 716, and 832\,nm. The angular resolutions of our images are between 30--40\,{\it mas}, and the limiting magnitude $V\sim16$\,mag. These speckle images are likely the highest possible angular resolution images of R136 in the optical as of this time until the arrival of the 30m class of telescope.  
Our imaging enables an improvement on currently available optical high angular resolution images of R136, and provides complementary images to those that may be taken in the near-infrared from either interferometry, or the James Webb Space Telescope.

This paper is organized as follows. Section 2 presents the observations and the data reduction, and Section 3 presents our results and the caveats associated with them. Conclusions are given in Section 4.

\section{Speckle imaging} \label{sec:style}

\subsection{Observations}

Zorro is the speckle instrument mounted on the 8.1m Gemini South telescope situated atop Cerro Pach{\'o}n, Chile \citep{scott21}. Light entering the instrument is split via a dichoric around 700\,nm, and falls on two EMCCDs, offering simultaneous speckle imaging in red and blue filters over an approximately 2.5$\arcsec\times2.5\arcsec$ field of view (FoV). 

A total of 40 1000-frame speckle data files having individual exposures of 60\,{\it ms} were acquired on R136 as a part of our project on 31 October 2021. The total integration time on source was 40\,minutes. The images were observed in the four medium-band filters EO466, EO562, EO716, and EO832{\footnote{https://www.gemini.edu/instrumentation/alopeke-zorro/}}, where the approximate central wavelength of the filter is denoted. The sky remained clear throughout the observations, with the full width half maximum (FWHM) measured using the P2 WaveFront Sensor hovering between 0.6--0.7$\arcsec$. Observations were conducted at a mean airmass of 1.3, with aim to observe the target as close as possible to the meridian to minimise atmospheric dispersion. Before the first file, after the last, and  at several points during the observations, a bright point source near on the sky was also observed, to provide estimates of the speckle transfer function throughout the observing sequence.
Three stars were used for this purpose, HR 1960, HR 1964, and HR 2221. The files taken on R136 were divided into four groups, with a point source before and after the group in each case. This was important to monitor and be able to correct for small changes in residual dispersion present due to the change of the zenith angle over the observational sequence. 

\begin{figure*}
\plotone{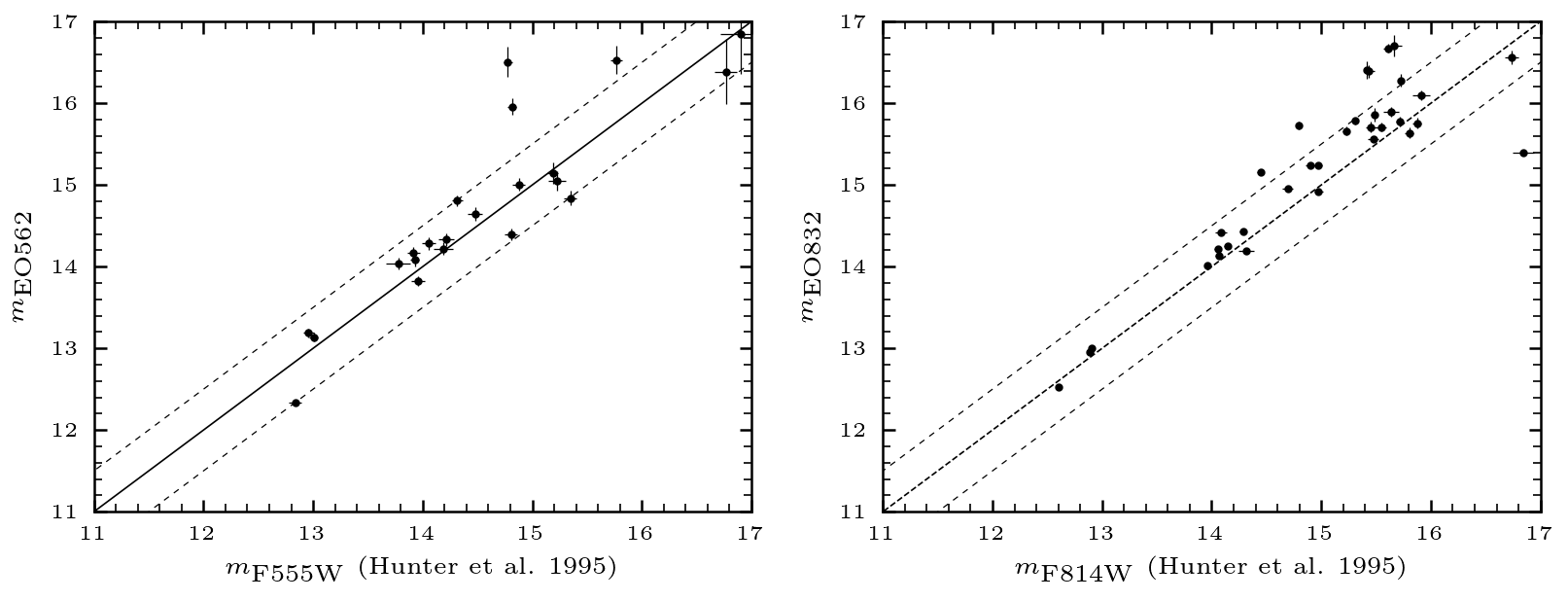}
\caption{Comparison of the Zorro photometry with the {\it HST} WFPC2 photometry from \cite{Hunter95}, for filters having similar central wavelengths. Solid lines represent unity, while dashed lines are for offsets of 0.1\,mag. Vega to AB offsets for the {\it HST} photometry were applied.  \label{magcomp}}
\end{figure*}

\subsection{Data reduction}

The speckle reduction employed began in the usual way, namely the computation of the  average autocorrelation functions of speckle frames of both R136 files and point source files. From this, the modulus of the Fourier transform of R136 can be obtained by Fourier transforming the autocorrelation of R136 to arrive at its power spectrum, dividing it by the average power spectrum of the point source calibrators, and taking the square root. 

What remains from that point in order to reconstruct the image of the target is the phase of R136 in the Fourier domain. However, the image morphology of the target is complex, and the phase function is difficult to reconstruct using normal speckle methods such as bispectral analysis. On the other hand, progress can be made by comparing the speckle data with infrared images of the same field taken in the infrared. The inverse-Fourier transform of the modulus shows peaks at the location of each point source relative to the brightest source in the field (which will be at the center of the field), as well as at the same separation from the central star but at a location corresponding to a position angle of 180$\degr$ from the actual location of each star. 

Our approach has been to identify peaks in common between the $H$-band image of our field (from \citealt{Camp10}, taken using the MAD instrument on the VLT) and the inverse-transform of the modulus obtained from our speckle data. We then assume the quadrant for each star identified is the same as in the infrared image. In some cases, the peak in the infrared image appears elongated and the modulus exhibits a double peak; we assume these are cases where there are two stars that were unresolved in the AO image but resolved in our speckle images due to the smaller diffraction limit in the visible range of the spectrum. In total, we were able to identify 41 peaks in common with the $H$-band image in images of the field taken taken through the EO832 filter, and 24 peaks in common when observing through the EO562 filter.

We next fitted the collection of peaks with Gaussian functions of various heights and positions on the image plane in the modulus. Only peaks with the quadrants that matched actual sources in the infrared image were fitted. The result was a list of positions and heights in the modulus which represented the actual star positions. To insure a robust fit, we fitted the three brightest stars in the field first using the power spectrum fitting method described in \cite{Horch96} and held those intensities fixed in our fit of the modulus. We then modeled how the height in the inverse-transformed modulus would correspond to intensity, and estimated the stellar intensities from this.
Using these, we could build an initial map of the star field, Fourier transform the result, and derive the phase function from that. Combining this phase function together with the original modulus, we in effect built the diffraction-limited Fourier transform of the reconstructed image. After low-pass filtering and inverse transforming, we arrive at a high SNR reconstructed image of the field. Thus, the AO images help to establish the quadrants of each source, but the intensity and location are determined from the speckle data. We completed this analysis for each group of observations on R136, as well as a separate analysis where all R136 files and all point source calibrator files were used in concert to arrive at a final result. Figure\,\ref{zorro} shows the final reconstructed  image obtained in the EO832 filter using all data files.

It is important to emphasize two caveats with the speckle image reconstructions made in this way. First, the method assumes that the image to be reconstructed is a collection of point sources and only uses peaks that are identifiable in the infrared AO images. Thus, the phase function used to create the reconstructed image is incomplete if other sources exist on the frame, or if some of the light detected cannot be represented as a pure point source. Second, the phase function is built from the fitted heights of the peaks in the inverse-transformed modulus. These may contain both random and systematic error originating from an imperfect point source match to the data or estimation of the intensities based on peaks in the inverse-transformed modulus. These effects need to be studied more fully, but what is clear at present is that if the phase function used to reconstruct the image is not correct due to an under- or over-estimated intensity, then it is possible to create false peaks in the reconstructed images that we calculate, at a level of at most a few percent relative to the maximum intensity in the image. Therefore, while the reconstructed images show much higher resolution than previously obtained of the field with AO, one should be careful not to assume that every peak in the reconstructed image comes from a stellar source at that location. However, some stars that appear blended in the IR images are clearly resolved here as seen from Figure\,\ref{comp}, and that first estimates of the visible-light magnitudes of many individual stellar sources in the field may derived from the speckle data.  

\subsection{Astrometry}

Precise astrometric solutions of the core of R136 are not currently available in the literature, due to the central density of the inner $1\arcsec$ of the cluster, and the angular resolution ($>1\arcsec$) of most wide-field surveys commonly utilised for astrometric calibration. The initial astrometric calibration was instead performed using the World Coordinate System (WCS) coordinates provided by Skiff (2016) {\footnote{https://vizier.cds.unistra.fr/viz-bin/VizieR?-source=J/ApJ/448/179}} for the \cite{Hunter95} {\it HST} images. An accurate transformation solution of the Zorro EO832 image coordinates to this WCS was obtained by correcting for distortion and isoplanatic effects using the IRAF task {\it msctpeak} in the tangent plate projection, combined with polynomials of order 4 (in the TNX WCS projection). Coordinates were estimated based on the pixel coordinates of the detected sources using the IRAF task {\it ccxymatch}, and the WCS calibrated image. Astrometric residuals are around 20{\it mas} or better. These coordinates were used to cross-match with the \cite{Hunter95} catalogue. Coordinates for all 41 sources in the red and 24 sources in the blue channel are taken from the EO832 image. They are listed in Table 1.
Cross-matches to the \cite{Crow16} and \cite{Khor21} were made on basis of the \cite{Hunter95} ID cited in those works. Note that the Skiff (2016) WCS has mean offsets around 0.15$\arcsec$ with respect to {\it Gaia} EDR3 astrometry. {\it Gaia} EDR3 coordinates cannot be used to obtain a useful WCS fit for our images, as they do not contain any astrometry for any star besides R136a1.

\begin{deluxetable*}{llccccccl}
\tablenum{1}
\tablecaption{Photometry and coordinates of sources detected by Zorro within the central core of R136.}
\tablewidth{0pt}
\tablehead{
  \colhead{\#} &
  \colhead{HSH95${^a}$} &
  \colhead{Right Ascension${^b}$} &
  \colhead{Declination} &
  \colhead{$m_{{\rm {EO466}}}$} &
  \colhead{$m_{{\rm {EO562}}}$} &
  \colhead{$m_{{\rm {EO716}}}$} &
  \colhead{$m_{{\rm {EO832}}}$} &
  \colhead{Spectral Type${^c}$}  }
\startdata
  0 & 3(a1) & 05$^h$38$^m$52$^s$398 & $-$69$\degr$06$\arcmin$02$\arcsec$86 & 12.192$\pm$0.02 & 12.334$\pm$0.03 & 12.541$\pm$0.02 & 12.522$\pm$0.02 & WN5h \\
  1 & 6(a3) & 05$^h$38$^m$52$^s$338 & $-$69$\degr$06$\arcmin$03$\arcsec$21 & 13.53$\pm$0.16 & 13.138$\pm$0.05 & 12.954$\pm$0.03 & 12.95$\pm$0.03 & WN5h \\
    2 & 5(a2) & 05$^h$38$^m$52$^s$417 & $-$69$\degr$06$\arcmin$02$\arcsec$81 & 13.397$\pm$0.08 & 13.187$\pm$0.06 & 12.94$\pm$0.03 & 12.998$\pm$0.02 & WN5h\\
  3 & 19(a6) & 05$^h$38$^m$52$^s$293 & $-$69$\degr$06$\arcmin$03$\arcsec$28 & 14.117$\pm$0.1 & 14.171$\pm$0.07 & 14.293$\pm$0.03 & 14.251$\pm$0.03 & O2If$^{d}$\\
  4 & 26 & 05$^h$38$^m$52$^s$298 & $-$69$\degr$06$\arcmin$03$\arcsec$34 & 14.237$\pm$0.21 & 14.219$\pm$0.08 & 14.097$\pm$0.03 & 14.135$\pm$0.03 \\
  5 & 21(a4) & 05$^h$38$^m$52$^s$348 & $-$69$\degr$06$\arcmin$02$\arcsec$54 & 13.922$\pm$0.15 & 13.825$\pm$0.06 & 14.235$\pm$0.03 & 14.221$\pm$0.02 & O2-O3\,V${^d}$\\
  6 & 27(a8) & 05$^h$38$^m$52$^s$364 & $-$69$\degr$06$\arcmin$02$\arcsec$53 & 14.332$\pm$0.13 & 14.335$\pm$0.08 & 14.462$\pm$0.03 & 14.426$\pm$0.03 & O2-O3\,V${^d}$\\
  7 & 24(a7) & 05$^h$38$^m$52$^s$416 & $-$69$\degr$06$\arcmin$02$\arcsec$50 & 14.03$\pm$0.32 & 14.282$\pm$0.08 & 14.39$\pm$0.04 & 14.415$\pm$0.03 & O3\,III(f*)\\
  8 & 20(a5) & 05$^h$38$^m$52$^s$438 & $-$69$\degr$06$\arcmin$02$\arcsec$66 & 14.095$\pm$0.18 & 14.085$\pm$0.08 & 14.001$\pm$0.03 & 14.016$\pm$0.02 & O2\,If*\\
  9 & 17 & 05$^h$38$^m$52$^s$390 & $-$69$\degr$06$\arcmin$02$\arcsec$79 & 13.68$\pm$0.46 & 14.037$\pm$0.08 & 14.079$\pm$0.04 & 14.191$\pm$0.03 & O5-O9\,V${^e}$\\
  10 & 35 & 05$^h$38$^m$52$^s$280 & $-$69$\degr$06$\arcmin$02$\arcsec$24 & 14.849$\pm$0.19 & 14.641$\pm$0.08 & 14.995$\pm$0.06 & 14.96$\pm$0.03 & O3\,V\\
  11 & 30 & 05$^h$38$^m$52$^s$341 & $-$69$\degr$06$\arcmin$02$\arcsec$32 & 14.646$\pm$0.22 & 14.808$\pm$0.07 & 15.145$\pm$0.07 & 15.161$\pm$0.04 & O7\,V\\
  12 & 70 & 05$^h$38$^m$52$^s$468 & $-$69$\degr$06$\arcmin$02$\arcsec$35 & 14.627$\pm$0.37 & 15.051$\pm$0.12 & 15.714$\pm$0.08 & 15.658$\pm$0.05 & O5\,Vz \\
  13 & 58 & 05$^h$38$^m$52$^s$491 & $-$69$\degr$06$\arcmin$02$\arcsec$55 & 14.31$\pm$0.34 & 15.008$\pm$0.07 & 14.94$\pm$0.05 & 14.919$\pm$0.03 & O3\,III(f*)\\
  14 & 50 & 05$^h$38$^m$52$^s$517 & $-$69$\degr$06$\arcmin$03$\arcsec$22 & 14.024$\pm$0.29 & 14.396$\pm$0.07 & 15.26$\pm$0.06 & 15.236$\pm$0.04 & O2-3\,V${^d}$ \\
  15 & 66 & 05$^h$38$^m$52$^s$432 & $-$69$\degr$06$\arcmin$03$\arcsec$29 & 15.013$\pm$0.28 & 15.147$\pm$0.13 & 15.788$\pm$0.07 & 15.782$\pm$0.04 & O3\,V\\
  16 & 62 & 05$^h$38$^m$52$^s$390 & $-$69$\degr$06$\arcmin$03$\arcsec$54 & 14.478$\pm$0.21 & 14.652$\pm$0.07 & 15.024$\pm$0.05 & 15.016$\pm$0.03 & O2-3\,V${^d}$ \\
  17 & 80 & 05$^h$38$^m$52$^s$345 & $-$69$\degr$06$\arcmin$03$\arcsec$66 & 15.108$\pm$0.34 & 14.841$\pm$0.09 & 15.782$\pm$0.06 & 15.706$\pm$0.05 & O8\,V\\
  18 & 52 & 05$^h$38$^m$52$^s$432 & $-$69$\degr$06$\arcmin$03$\arcsec$94 & 15.673$\pm$0.39 & 15.959$\pm$0.1 & 15.279$\pm$0.07 & 15.238$\pm$0.03 & O3\,V\\
  19 & 48 & 05$^h$38$^m$52$^s$621 & $-$69$\degr$06$\arcmin$02$\arcsec$79 & 16.005$\pm$0.23 & 16.506$\pm$0.18 & 15.804$\pm$0.12 & 15.728$\pm$0.04 & O2-3\,III(f*)\\
  20 & 112 & 05$^h$38$^m$52$^s$538 & $-$69$\degr$06$\arcmin$02$\arcsec$16 & 15.916$\pm$0.28 & 16.531$\pm$0.17 & 15.812$\pm$0.1 & 15.757$\pm$0.06 & O8.5\,III(f)\\
  21 & 257 & 05$^h$38$^m$52$^s$320 & $-$69$\degr$06$\arcmin$02$\arcsec$38 & 16.391$\pm$0.22 & 16.853$\pm$0.49 & 16.054$\pm$0.06 & 16.015$\pm$0.05 \\
  22 &  & 05$^h$38$^m$52$^s$261 & $-$69$\degr$06$\arcmin$02$\arcsec$29 & 16.177$\pm$0.41 & 17.228$\pm$0.18 & 16.031$\pm$0.33 & 15.941$\pm$0.1 \\
  23 & 231 & 05$^h$38$^m$52$^s$522 & $-$69$\degr$06$\arcmin$02$\arcsec$39 & 16.481$\pm$0.51 & 16.388$\pm$0.4 & 15.432$\pm$0.17 & 15.399$\pm$0.03 \\
  24 & 78 & 05$^h$38$^m$52$^s$219 & $-$69$\degr$06$\arcmin$02$\arcsec$96 &  &  & 15.556$\pm$0.08 & 15.558$\pm$0.03 &O4:\,V \\
  25 & 90 & 05$^h$38$^m$52$^s$257 & $-$69$\degr$06$\arcmin$02$\arcsec$72 &  &  & 16.067$\pm$0.09 & 15.775$\pm$0.06 & O5\,V \\
  26 & 92 & 05$^h$38$^m$52$^s$277 & $-$69$\degr$06$\arcmin$02$\arcsec$03 &  &  & 15.587$\pm$0.09 & 15.639$\pm$0.06 & O3\,V \\
  27 & 93 & 05$^h$38$^m$52$^s$412 & $-$69$\degr$06$\arcmin$02$\arcsec$39 &  &  & 16.717$\pm$0.19 & 16.707$\pm$0.13 & \\
  28 & 89 & 05$^h$38$^m$52$^s$503 & $-$69$\degr$06$\arcmin$03$\arcsec$12 &  &  & 15.888$\pm$0.09 & 15.897$\pm$0.06 & O4\,V$^{d}$\\
  29 &  & 05$^h$38$^m$52$^s$344 & $-$69$\degr$06$\arcmin$03$\arcsec$61 &  &  & 17.387$\pm$0.16 & 17.154$\pm$0.15 & \\
  30 & 75 & 05$^h$38$^m$52$^s$200 & $-$69$\degr$06$\arcmin$01.70 &  &  & 15.726$\pm$0.09 & 15.859$\pm$0.09 & O4\,V$^{d}$ \\
  31 & 73 & 05$^h$38$^m$52$^s$558 & $-$69$\degr$06$\arcmin$03$\arcsec$37 &  &  & 16.403$\pm$0.25 & 16.391$\pm$0.08 & O9\,V \\
  32 & 69 & 05$^h$38$^m$52$^s$549 & $-$69$\degr$06$\arcmin$03$\arcsec$54 &  &  & 15.723$\pm$0.1 & 15.71$\pm$0.07 & O3-6\,V \\
  33 & 94 & 05$^h$38$^m$52$^s$545 & $-$69$\degr$06$\arcmin$03$\arcsec$88 &  &  & 16.261$\pm$0.15 & 16.282$\pm$0.08 & O3\,V \\
  34 &  & 05$^h$38$^m$52$^s$349 & $-$69$\degr$06$\arcmin$03$\arcsec$25 &  &  & 18.078$\pm$1.1 & 18.031$\pm$0.1 & \\
  35 & 86 & 05$^h$38$^m$52$^s$326 & $-$69$\degr$06$\arcmin$03$\arcsec$06 &  &  & 16.649$\pm$0.1 & 16.674$\pm$0.05 & O3-4\,V$^{d}$ \\
  36 & 108 & 05$^h$38$^m$52$^s$181 & $-$69$\degr$06$\arcmin$01.78 &  &  & 16.267$\pm$0.09 & 16.102$\pm$0.06 & O7-8\,V$^{d}$\\
  37 &  & 05$^h$38$^m$52$^s$174 & $-$69$\degr$06$\arcmin$01.72 &  &  & 16.92$\pm$0.17 & 16.922$\pm$0.06 & \\
  38 &  & 05$^h$38$^m$52$^s$176 & $-$69$\degr$06$\arcmin$02$\arcsec$98 &  &  & 16.378$\pm$0.1 & 16.291$\pm$0.07 & \\
  39 & 203 & 05$^h$38$^m$52$^s$178 & $-$69$\degr$06$\arcmin$03$\arcsec$05 &  &  & 16.577$\pm$0.08 & 16.557$\pm$0.08 & \\
  40 & 77 & 05$^h$38$^m$52$^s$478 & $-$69$\degr$06$\arcmin$04$\arcsec$03 &  &  & 16.391$\pm$0.35 & 16.409$\pm$0.11 & O5.5\,V+O5.5\,V \\
\enddata
\tablecomments{(a) HSH95 refers to the nomenclature from \cite{Hunter95}, with the nomenclature in the parenthesis is for R136\,a sources from \cite{Baier}. (b) Coordinates are in J2000 epoch, with the reference World Coordinate System from Skiff (2016). (c) Spectral types are optical spectral types from \cite{Crow16}, or otherwise indicated. (d) Ultraviolet spectral type from \cite{Crow16}. (e) Photometric spectral type based on Zorro magnitudes. }
\end{deluxetable*}

\subsection{Photometry}
Reconstructed speckle images always place the brightest star at the center, with the {\it psf} normalised to 1, only providing relative photometry to the brightest source. To obtain an absolute calibration, we observed at the same median airmass of our science observations (1.33), speckle images in the same filter sets of the spectrophotometric standard EG\,21. Using the speckle reduction, the number of counts detected per frame is computed, which is transformed to the count rate per second. This provides instrumental magnitudes from the summed frames in each filter. From the flux calibrated spectra of EG21, AB magnitudes were calculated for each filter, by convolution of the spectra with the filter, dichroic, and detector quantum efficiency curves. This yielded instrumental zero-points, from which we calculate AB magnitudes, using the percentage of observed counts contributed by the central star, to estimate the correction to the central star magnitude, and place all the photometry on an AB magnitude scale. 
\begin{figure}
\plotone{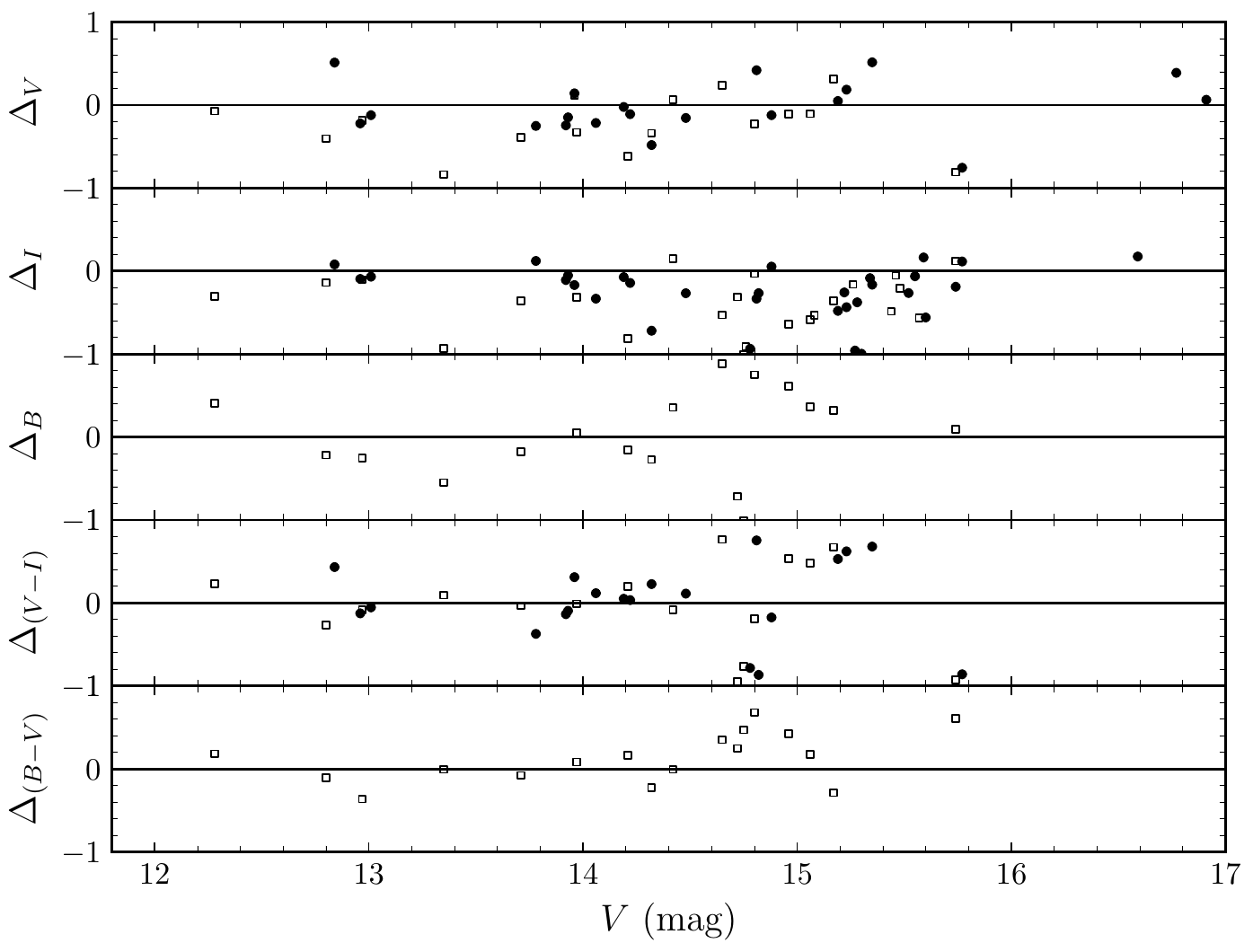}
\caption{Residuals between {\it HST} and Zorro photometry, where {\it B,V} and {\it I} represent the {\it HST} F438W, F555W, F814W, and Zorro EO466, EO532, and EO832 filters, respectively. Circles correspond to residuals of WFPC2 photometry from \cite{Hunter95}, while squares are the comparisons with \cite{Crow16} WFC3 photometry. Vega to AB offsets for the {\it HST} photometry were applied. \label{magresd}}
\end{figure}

\cite{Horch01,Horch04} showed that there are systematic
errors in the differential magnitudes in speckle imaging due to the finite size of the isoplanatic angle. This results in fainter tertiary stars as the distance from the central star increases; because the light no longer travels through the same air column to reach the telescope leading due a different turbulence pattern at the wave front. This degree of correlation between the magnitude of the primary star and remaining components is given by the $q$ parameter. Here, $q=\rho\omega$, where $\rho$ is the distance from the central star in arcseconds, and $\omega$ the seeing in arcseconds. To calculate the magnitude difference, we used the results from \citep{How19} for the EO\,832 filter. A correction of $0.2802\times q$ was applied to the magnitudes, with total magnitude corrections ranging from 0.1--0.4\,mag. From the results in \citep{Horch04}, there appear no discernible difference based on central wavelength of the observations for similar speckle imagers, so no filter based correction was applied. 

The final calibrated magnitudes and astrometry of all sources are listed in Table 1. Photometric uncertainties include the zero-point errors propagated with the measured differential magnitude uncertainties.

\subsection{Comparisons with literature photometry}

\cite{Hunter95} presented the first resolved optical photometry of R136 using the WFPC2 camera on\,board the {\it HST}, using short exposures of 4, 5, and 10 seconds in the F336W, F555W, F814W (hereafter $U$, $V$, and $I$) filters respectively. Deeper photometry using the advanced UVIS1 chip on the WFC3 camera was presented in the F336W, F438W (hereafter $B$) F555W, and F814W by \citep{Crow16}.

We compared the photometry with the Zorro filters having similar central wavelengths. While {\it HST} photometry is usually converted using color equations including a color term, we do not employ such equations in our comparisons. This is because color terms of the Zorro medium-band filters, on examination using stellar atmosphere models of hot stars do not straightforwardly compare with colors of the broad-band {\it HST} filters. Converting the magnitudes between the medium-band Zorro and broad-band {\it HST} filters using color equations derived from model atmospheres, without a significant sample of observed comparable photometry in each filter thus adds significantly to the noise in comparisons, rather than reduce it. Instead, we ignore color-based terms and compare directly the {\it HST} $B$, $V$, and $I$ magnitudes with Zorro $m_{\rm EO466}$, $m_{\rm EO562}$, and $m_{\rm EO832}$ magnitudes respectively, only applying Vega to AB magnitude system conversion factors for the {\it HST} magnitudes. The comparisons to the WFPC2 photometry is shown in Figure\,\ref{magcomp}, with the residuals from the comparisons shown in Figure\,\ref{magresd}.

The photometry from \cite{Hunter95} has 35 stars in common, and compares favorably against our photometry. The median offset for ($V-m_{\rm EO562}$) is $-$0.12\,mag with a variance of 0.27, for ($I-m_{\rm EO832}$) is $-$0.09 with a much smaller variance of 0.07. This is consistent with the expectation from hot star standard spectra between these two systems, having median offsets of $-0.05$ and $-0.06$, while not accounting for any color terms. We draw attention to the three brightest WN5h stars, R136 a1, a2, and a3. For a1, the \cite{Hunter95} $V$ magnitude is 0.5\,mag fainter that our photometry, while for the other two stars our photometry is around 0.1-0.2 magnitudes fainter, more consistent with expectations. A similar trend is found for the comparison the $I$-band photometry, although the differences are of the order of 0.05\,mag. 

A comparison to the \citep{Crow16} photometry presents larger variance, for a sample of 24 stars. We find for a comparison with the $B$, $V$, and $I$ magnitudes median offsets of $-0.21$, $-0.2$, and $-0.34$ respectively. This is larger than expectations using standard star spectra, and similar in scale to the offsets found by \citep{Crow16} against \cite{Hunter95} photometry. Our results, and those from \cite{Hunter95} are fainter for the brightest stars those presented in \cite{Crow16}. Finally, the variance for ($B-m_{\rm EO466}$) is 0.3, and displays the comparatively larger deviations in the bluest photometry for the faintest stars as seen in Figure\,\ref{magresd}.

\section{Results}

\subsection{Stellar census}

\begin{figure}
\plotone{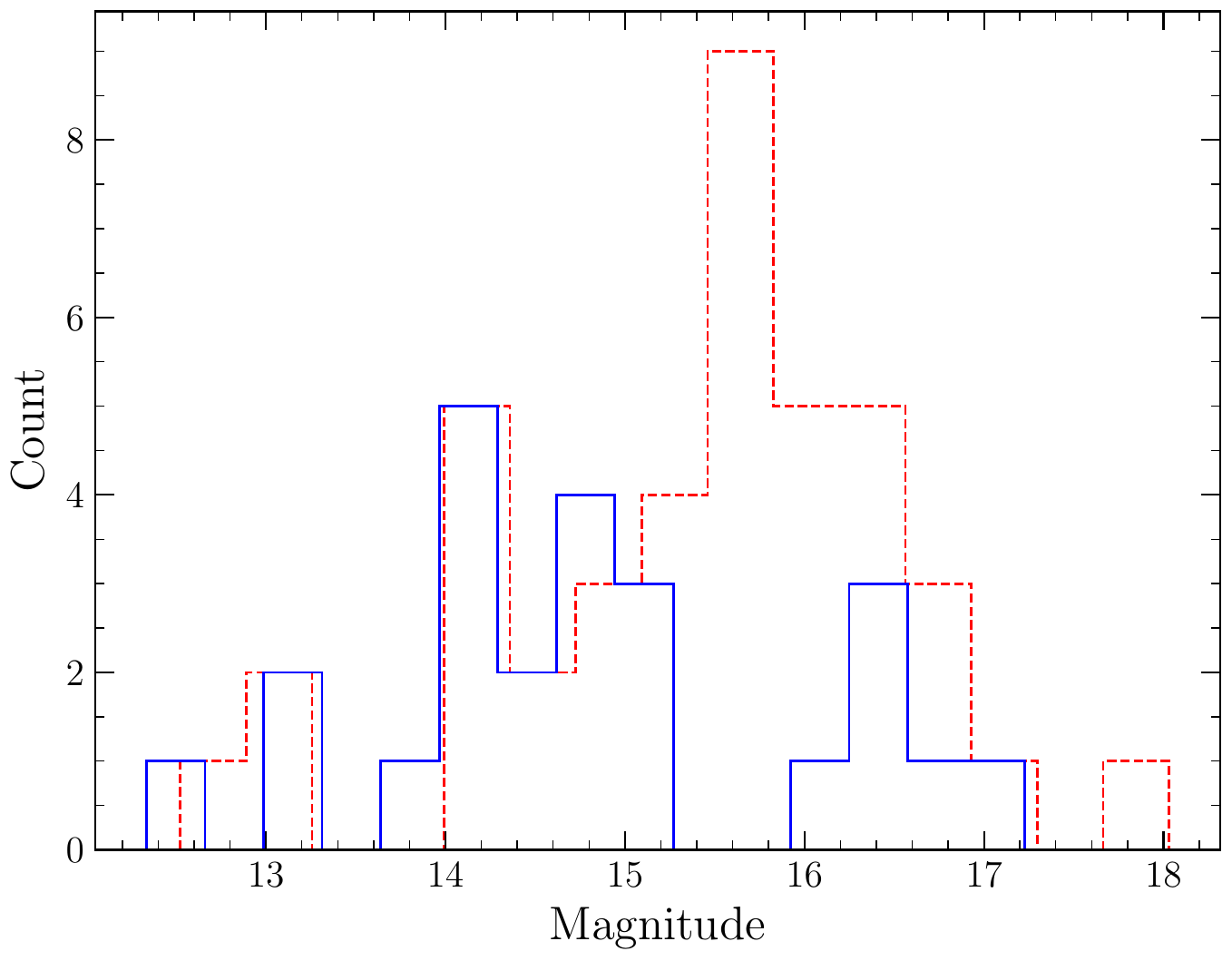}
\caption{Observed luminosity function from the Zorro photometry in $m_{{\rm {EO832}}}$ (red dashed line), and m$_{{\rm {EO562}}}$ (blue solid line) filters, representing the sources recovered in the red and blue channels respectively.     \label{histfig}}
\end{figure}

\begin{figure}
\plotone{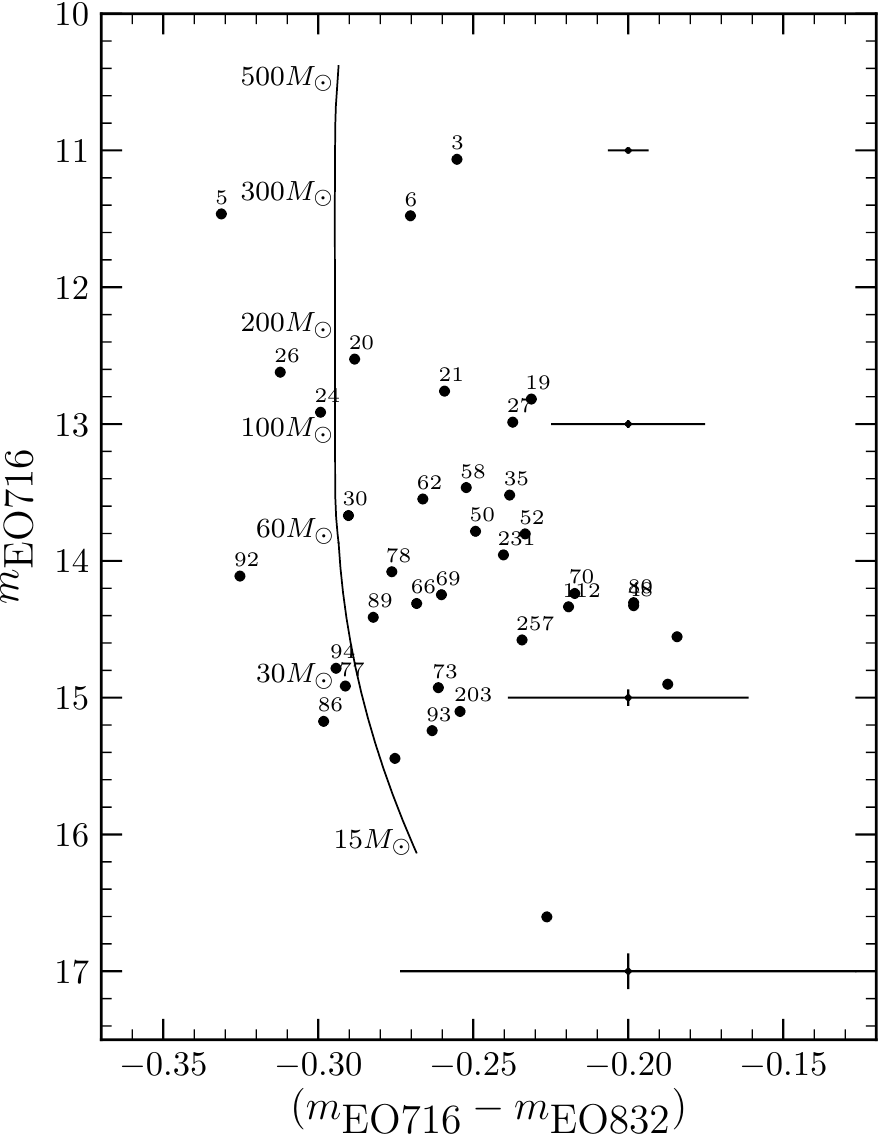}
\caption{Color-magnitude diagram of ($m_{{\rm {EO716}}}$-$m_{{\rm {EO832}}}$) vs. $m_{{\rm {EO716}}}$ for all sources. The \cite{Hunter95} identification numbers are written when there is a cross-match. Smoothed zero-age main sequence isochrone from 60--500\,$M_{\odot}$ from \cite{Kohler}, and from 15--60\,$M_{\odot}$ from \cite{Brott} are shown, converted to the observational plane using the stellar atmosphere models of \cite{tlusty}. The stellar mass at various $m_{{\rm {EO716}}}$ are shown, along with median errors at different steps of $m_{{\rm {EO716}}}$. The ZAMS is corrected for a distance modulus of 18.48\,mag, and assumes a mean extinction of $A_V=2$\,mag, having $R_V$=4.18 following the reddening law of \cite{jma13}.      \label{cmdfig}}
\end{figure}

Table 1 presents the photometry of all sources identified. 41 sources are identified in the red filters, and 24 in the blue. In Figure\,\ref{histfig} the $m_{{\rm {EO832}}}$ and $m_{\textrm EO555}$ luminosity function of all 41 (24) sources are respectively presented. From the magnitude histogram in Figure\,\ref{histfig}, the 90\% completeness limit in the red channel corresponds to $m_{{\rm {EO832}}}=$15.75\,mag, which is close to the sensitivity limits of Zorro{\footnote{https://www.gemini.edu/instrumentation/alopeke-zorro/capability\#Sensitivity}}. In the blue channel, the number of sources recovered is much lower. However, all stars brighter than $m_{{\rm {EO832}}}<15.5$\,mag were detected in both the blue and the red channel. Given that a magnitude of $V\sim15$ corresponds to an late O/early B type star at the adopted distance, assuming an absolute extinction of $A_V=2$\,mag ({\citealt{Crow16} suggest 1.72$\pm$0.25\,mag), we should recover all known and visible O type stars within our FoV in all the observed passbands. 

\begin{deluxetable*}{rlllll}
\tablenum{2}
\tablecaption{Flux ratios of stars within the central core from different photomeric studies.}
\tablewidth{0pt}
\tablehead{
  \colhead{Instrument} &
  \colhead{Zorro} &
  \colhead{WFPC2${^a}$} &
  \colhead{SPHERE${^b}$} &
  \colhead{WFC3${^c}$} &
  \colhead{AO Spectroscopy${^c}$} \\
\colhead{}&\colhead{466/562/716/832} & \colhead{F336W/F555W/F814W} & \colhead{H/Ks} & \colhead{F336W/F555W/F814} & \colhead{1500A/UV/Ks}
}
 \startdata
  a2/a1 & 0.33/0.46/0.69/0.65 & 0.70/0.90/0.76 & 0.69/0.77 & 0.88/0.95/0.84 & 0.56/0.62/0.76\\
  a3/a1 & 0.29/0.48/0.68/0.67 & 0.76/0.86/0.77 & 0.67/0.76 & - & -\\
  HSH17$^{d}$/a1 & 0.25/0.20/0.24/0.22 & 0.27/0.42/0.21 & 0.13/0.13 & - & -\\
\enddata
\tablecomments{(a) Taken from optical {\it HST} WFPC2 data of \cite{Hunter95} (b) From IR AO imaging of \cite{Khor21} (c) Based on {\it HST} WFP3 in \cite{Crow10}, $\S$2.3. (d) HSH17 is the source nomenclature taken from \cite{Hunter95}, and refers to the marked companion to a1 in Figure\,1. }
\end{deluxetable*}

We compared our census to the stars detected within the FoV by \cite{Hunter95} and \cite{Crow16}, with magnitude $V<16$\,mag. Compared to \cite{Hunter95}, all stars were identified in the images. However, the magnitudes of HSH\,107 and HSH\,119 could not be recovered (the ID denoted is from \citealt{Hunter95} here), as they are relatively faint stars ($V>15.5$\,mag) near bright stars. Their photometry is not presented in our results. All stars from the WFC3 data of \cite{Crow16} were recovered. 

New sources within this region were identified by our photometry. We draw attention to R136a3, which has a newly identified visual component within $\lesssim$2000\,au, previously unidentified (\#34). This star is the faintest in the sample, having $m_{{\rm {EO832}}}$ of 18\,mag. In addition, R136\,a6 is separated into two components cleanly for the first time in the optical, as well as separating the central a1-a2 system from the more fainter HSH\,17 (\#9 in Table\,1), located $\sim$3000\,au (70\,mas) from a1.

The color-magnitude diagram (CMD) of ($m_{{\rm {EO716}}}-m_{{\rm {EO832}}}$) vs. $m_{{\rm {EO716}}}$ is shown in Figure\,\ref{cmdfig}. Overlaid is the zero-age main sequence (ZAMS) locus of rotating massive stars having initial rotational velocities of 150\,$km\,s^{-1}$. For masses between 60--500\,$M_{\odot}$, the locus from \cite{Kohler} is shown, and for masses between 15--60\,$M_{\odot}$ the \cite{Brott} ZAMS is adopted. The models are transformed to the observational plane using the 0.5$Z_{\odot}$ stellar atmosphere models from \cite{tlusty} and the appropriate instrumental responses. 

All sources are above $15\,M_{\odot}$ following the ZAMS models, except for \#34 (visual companion to R136\,a3), with the maximum around 300\,$M_{\odot}$ for the three WN5h stars. The most massive stars after the WN5h stars are the early O supergiants found by \cite{Crow16}. These include \#8 (O2If*), \#7 (O3IIIf*), \#5 (O2-O3V), \#6 (O2-O3V), \#3 (O2If), and \#4, having estimated masses between 70--150\,$M_{\odot}$ \citep{Crow16}. Numbering corresponds to Table 1. There exists a gap in known stellar masses between the early O supergiants, and the WN5h stars in this region.  

\begin{figure*}
\plotone{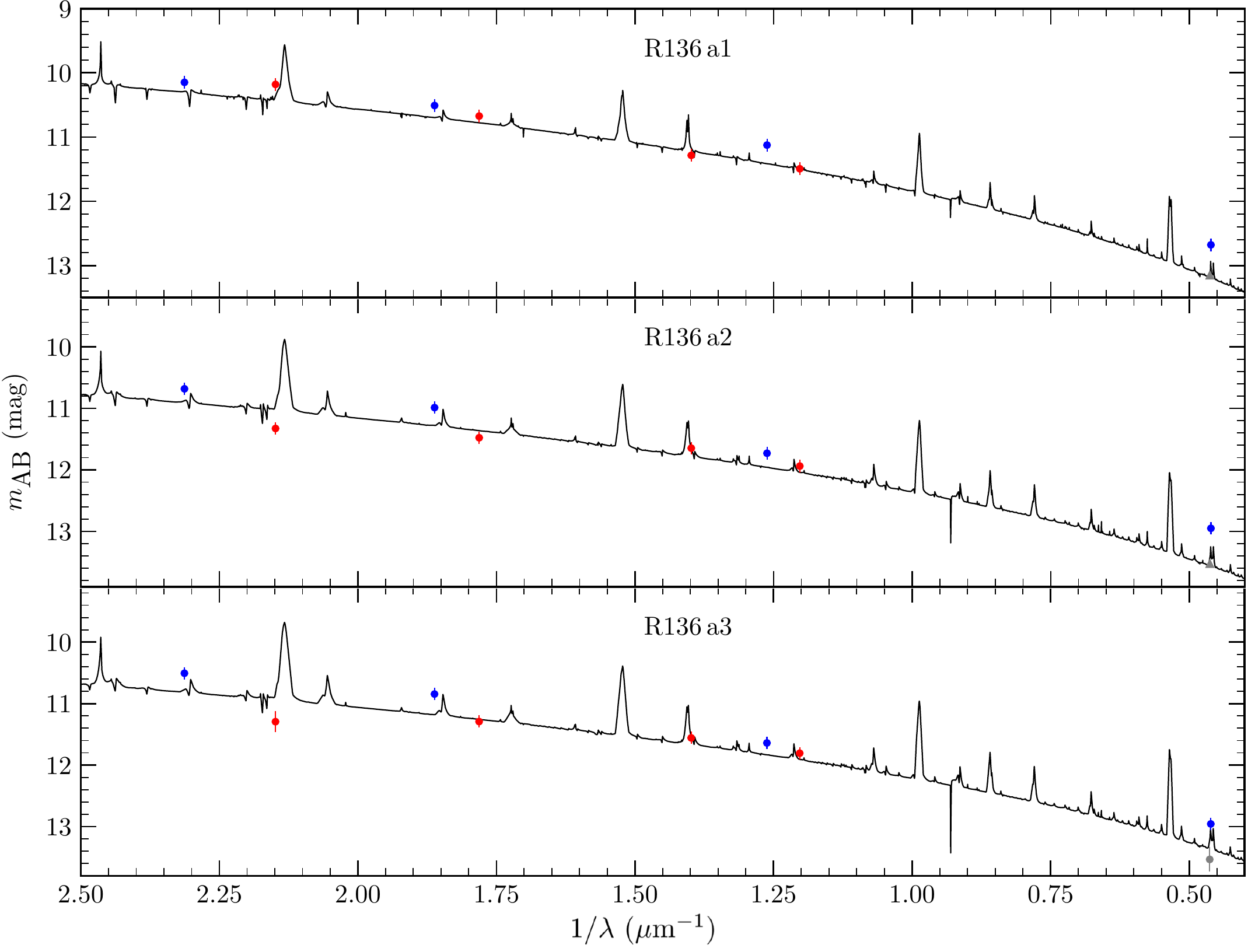}
\caption{Spectral energy distribution of the three WN stars in R136, labeled. Underlying theoretical models are from \cite{Best20}. Observed Zorro photometry are given by red circles, with the blue circles representing {\it HST} or $K$s photometry from \cite{Khor21}, plotted in AB magnitude units. Gray circles or carets (lower limits) are $K$ magnitudes from \cite{Rubio}. All error bars are of 0.1\,mag, except when greater. }    \label{seds}
\end{figure*}

\subsection{Core of R136}

The crowding within the central 0.1$\arcsec$ of R136 is thought to have led to differences between various photometry presented in the literature. Table 2 lists the flux ratios of the a1/a2/HSH17 system at the center of R136, and also with the other WN5h star, a3. HSH17 is the visual companion to a1, marked by the arrow in Figure\,1, and the nomenclature taken from \cite{Hunter95}. The flux ratios of a1/a2 in Zorro, AO IR, and AO spectroscopy is much lower than from WFC3 photometry indicating possible saturation in the WFC3 photometry (the magnitudes of the a1/a2 system in \citealt{Crow16} are corrected to give a flux ratio of 0.62). WFPC2 photometry gives a higher flux ratio in the $V$ filter, but lower in $U$ and $I$. a1 appears significantly brighter than a2, with a median flux ratio at $m_{\textrm{EO832}}$ or $m_{\textrm{EO716}}$ of 0.65 and 0.69 respectively, but at $m_{\textrm{EO466}}$ around 0.33. This is closer to the flux ratio seen in 1500\AA\ spectroscopy (0.56) reported in \cite{Crow16}. In principle, the observed flux ratios should be similar across similar wavelengths under the reasonable assumptions of similar underlying spectral type and extinction. The difference in flux ratios suggests the possibility of either lower extinction towards a1 (as a1 has excess UV flux compared to a2 assuming both have similar WN5h spectral types), or significant emission line flux. Future resolved optical spectroscopy of the He{\scriptsize II}\,$\lambda$4686\AA\ line in the a1/a2 system can investigate this issue.

While HSH17 is not described in the WFC3 photometry presented in \cite{Crow16}, in WFPC2 photometry it has a flux ratio agreeing with our Zorro estimates. It's brightness is similar to the known O spectral types, and it is located $\sim$3000 au away from a1. To determine its spectral type, we fit the available Zorro photometry along with IR AO $K$s photometry from \cite{Khor21} to \cite{tlusty} stellar atmosphere models. It is brighter in $m_{\textrm{EO466}}$ (comparable to a3 and a2) than the redder bandpasses, suggesting significant He{\scriptsize II} $\lambda$4686\AA\ emission. The resulting best fit is poor ($\chi^2 \sim$80), but suggests an mid O dwarf spectral type (having effective temperature of 35$\pm$10kK), having absolute extinction of 1.65\,mag, with the $m_{\textrm{EO466}}$ magnitude fit as an upper limit due to possible contamination from He{\scriptsize II} $\lambda$4686\AA\ emission line flux. Given its extremely blue location in the color-magnitude diagram (Figure\,\ref{cmdfig}), and probable He{\scriptsize II} $\lambda$4686\AA\ emission the star itself could be considerably hotter with it's spectral type not well represented by \cite{tlusty} model atmospheres. 

We also consider the possibility that the detected visual binaries are not chance alignments with R136a1, and a3. HSH17 is $\sim$70\,mas away from R136a1, and the visual binary of of R136a3 is $\lesssim$40\,mas from it. To calculate the probability of chance alignments we assume a cluster surface density of $\sim$20\,$arcsec^{-2}$ around the central 1$''$ of R136a1 (from \citealt{Hunter95}), and use the measured angular separations, adopting the formulation of \cite{binary}. Our calculation suggests that there is $\sim$75\% chance that HSH17 is not a chance alignment with R136a1, and that \#34 is around $\sim$90\% not aligned by chance with a3. This opens up statistically, the non-negligible probability that these stars are actually part of a binary system.

\subsection{Masses of the WN5h stars}

Spectral energy distribution (SED) models of the central WN5h stars, R136a1, a2, and a3 are presented in \cite{Best20}. The spectral parameters of the SEDs were derived by comparison to available spectroscopy. The authors of that paper estimated the stellar luminosity by fitting the SEDs against available photometry from \cite{Crow16} and \cite{Khor21}, accounting for extinction. 

To derive the luminosity using Zorro magnitudes, we compared our available photometry against the SEDs, assuming the reddening law from \cite{jma13} having the mean $R_V=4.18$ of the region \citep{Best20}\footnote{The extinction coefficients computed following \cite{jma13} are $A_{{\rm {EO466}}}/A_{{V}}$=1.18, $A_{{\rm {EO532}}}/A_{{V}}$=0.98, $A_{{\rm {EO716}}}/A_{{V}}$=0.74, and $A_{{\rm {EO832}}}/A_{{V}}$=0.6. $A_V$ designates the monochromatic extinction at 5495\,\AA\ }. The value of absolute extinction was left free between the range of 1.6$<A_V<$2.0, with the total luminosity the remaining free parameter. The extinction coefficients for each filter were calculated using the integrated response assuming the \cite{jma13} reddening law for the value $R_V=4.18$. No Milky Way, or foreground extinction was included in the fitting process, following the prescription in \cite{jma13}. A simple least square fit was performed searching for the $\chi^2$ minimum for the extinction and total luminosity value. The total integrated luminosity is calculated based on the resulting fit. 

The resulting fits are plotted in Figure\,\ref{seds}. Also shown are the archival {\it HST} photometry from \cite{Hunter95}, and $K$s magnitudes from \cite{Khor21}. All magnitudes are shown with error bars of 0.1\,mag for illustrative purposes, although the fitting was performed with the error bars listed in Table 1. The error bars are chosen such because a magnitude uncertainty of 0.2\,mag leads to approximately an error in the logarithm of luminosity in units of solar luminosity of $\sim$0.1\,dex. Also shown are the $K$s-band magnitudes derived from integral field unit spectroscopy from the SINFONI instrument on the Very Large Telescope estimated by \cite{Rubio}. The optical {\it HST} and IR AO photometry predict smaller values of magnitude (and thereby higher luminosity) than our magnitude estimates. A considerable fraction of this for the central sources could be due to contamination from the neighboring visual companions, which are not separated well. The $K$s magnitudes from \cite{Rubio} agree better with our results, and might do so due to the background subtraction methodology adopted by those authors. 
\begin{figure}
\plotone{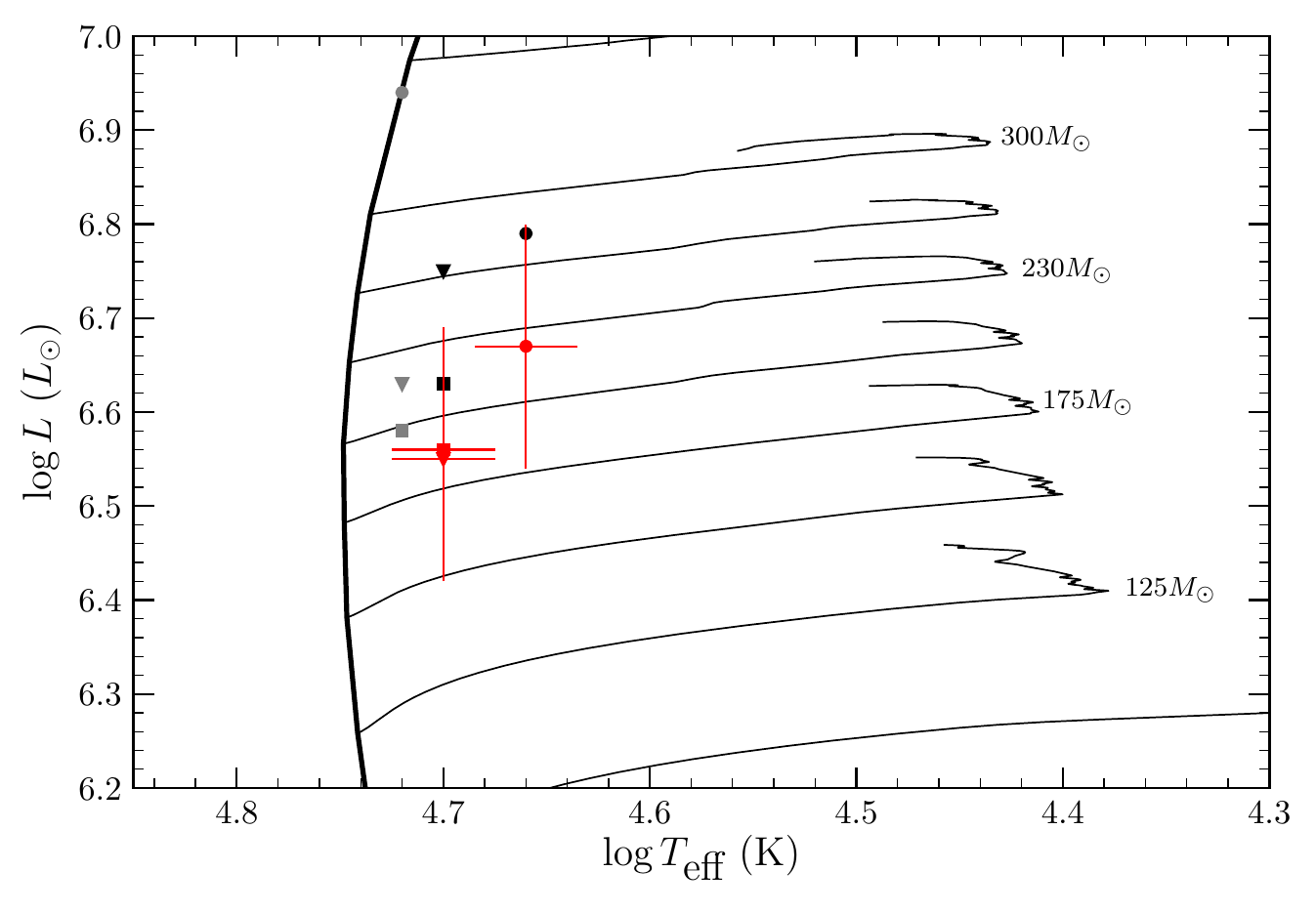}
\caption{Hertzsprung-Russell diagram of the WN5h stars in R136. R136a1 is marked as a circle, a2 as a caret, and a3 a square, while red represents results in this work, black for values from \cite{Best20}, and gray the results of \cite{Crow16}. The stellar mass tracks are taken from \cite{Kohler}, for the initial rotational velocity of 150\,$km\,s^{-1}$.} 
\label{hrd}
\end{figure}

The resulting values of luminosity are given in Table\,3. From the derived luminosity values, and adopting the effective temperatures from \cite{Best20}, we estimate stellar initial masses using BONN Stellar Astrophysics Interface (BONNSAI; \citealt{bonnsai}). BONNSAI provides Bayesian probability distributions of the fundamental stellar parameters (such as initial stellar mass, current mass, age) by comparing observed stellar parameters against stellar evolutionary models, while providing uncertainties on the estimated parameters, and whether they disagree with current models. The stellar evolutionary models used in BONNSAI are from \cite{Kohler}. To estimate the stellar initial mass, we used our estimated luminosity and its associated uncertainty, and the effective temperature, stellar mass loss rate, rotation velocity, and helium abundances from \cite{Best20}. The positions of the WN5h stars are shown in the Hertzsprung-Russell diagram in Figure\,\ref{hrd}, and the resulting initial masses, as well as a comparison with those available in the literature are given in Table\,3. The resulting luminosities, and initial masses are comparatively smaller than estimated previously by \cite{Crow16} and \cite{Best20}, and similar to those estimated in \cite{Rubio}.

Although there is no theoretical upper-mass limit to the initial mass function, the empirical upper-mass limit was considered to be around 150\,$M_{\odot}$ \citep{figer, kroupa}. In 2010, the three central stars of R136 were argued by \cite{Crow10} to have initial masses greater than $\sim$150\,$M_{\odot}$ (although it has been argued in \citealt{pavel1, pavel2, pavel3} that these VMS stars may have formed from mergers that could occur frequently in such binary-rich stellar-dynamical environments). The empirical stellar upper-mass limit remains set by R136\,a1, whose current mass estimates in the literature are between 250--325\,$M_{\odot}$ \citep{Crow16, Best20}. Based on luminosities determined from the resolved Zorro imaging of a1, the the mass we determine is considerably lower at $196^{+34}_{-27}\,M_{\odot}$, pointing towards a lower upper-mass limit than previously considered, and similar to the estimates obtained by \cite{Rubio}.

The key relevance of the increased upper-mass limit estimate from 150 to 300 $M_{\odot}$ found by \cite{Crow10} (under the assumption of non-binarity) was that the observed mass range comfortably encompasses the hypothesized pair-instability supernova (PISN) range between 140-260 $M_{\odot}$ \citep{heger}. Note that these mass ranges were derived for zero metallicity, as finite metallicity stars are expected to lose more mass in radiation-driven winds. As the mass-loss rates for very massive stars are likely underestimated, enhanced mass-loss rates would allow for a more substantial mass-loss history, and potentially a higher upper-mass limit \citep{vink}. Nonetheless, the present day mass estimates of the most massive stars in conjunction with hydrogen abundances provided key constraints on the evolutionary mass history and therefore the actual upper-mass limit \citep{Crow10}.

Therefore, given that stars around 300 $M_{\odot}$ were considered to exist even in the Local Universe had huge implications for stellar evolution studies -- leading to increased numbers of black holes and gravitational wave events, and various feedback mechanisms, such as ionizing radiation which are strong functions of stellar mass \citep{schneider}. Most dramatically, even if stars in relatively high-$Z$ environments such as the LMC ($\sim$\,1/2\,$Z_{\odot}$), lost too much mass, the very existence of stars around 300 $M_{\odot}$ led to the logical conclusion that PISNe should exist, even in the Local Universe \citep{Crow10}. The importance of whether or not PISNe exist cannot be overemphasized, as just 1 PISN from a 300 $M_{\odot}$ star would produce and release more metals into the interstellar medium (ISM) than an entire stellar mass function below it \citep{langer}, which would completely change our understanding of galactic chemical evolution modeling.

But despite a decade of searching for evidence of PISN in super-luminous SNe observations, observers have yet to find a reliable PISN candidate. Furthermore, the PISN theory predicts a conspicuous yield pattern with abundances of even $Z$ elements consistently higher than those of odd $Z$ elements (the so-called ``odd-even'' effect), but observations of carbon-rich chemically extremely metal- poor (CEMP) stars have thus far nor revealed these specific signatures \citep{umeda}. The implication of our lowered 200 $M_{\odot}$ mass for R136a1 is that the stellar upper-mass limit is lower, naturally explaining that PISNe are avoided, and accounts for the lack of very long-lived light-curves amongst super-luminous SNe, as well as the absence of the odd-even signature in CEMP stars.

There are two important caveats to our result. Firstly, the stellar luminosity considered here for determination of the mass would be inaccurate in the case of a multiple system. From our observations, we are only able to separate visual binary components farther than $\sim$2000\,au apart. \cite{Crow10} consider the case of multiples, and ruled out that the three central WN5h stars were equal mass close binaries ($<$200\,au) based on the X-ray luminosity, which would be higher in case of colliding winds. Similarly, \cite{schnurr} carried out a spectroscopic study of R136, and determined that none of the central stars have binaries with orbital periods shorter than 44\,{\it days}. If we assume a modest mass ratio of 0.25, this would rule out binaries closer than $\sim$2\,au The parameter space between radial velocity observations, and our speckle imaging, can be filled by the advent of 30m class of telescopes, which at diffraction limit can reach angular resolutions of $\sim$200\,au. 

The second caveat is the dependency on the evolutionary models used to estimate the stellar mass. At such high stellar masses, a significant fraction of the star's observed mass is lost before it becomes visible. A suitable treatment of the mass loss due to radiation-driven winds must be accounted for. The models used in \cite{bonnsai} may have mass-loss rates underestimated by up to a factor of about two \citep{Best20}, which could lead to a much higher initial mass estimate.  \cite{vink} use radiative transfer models to show that an exact mass-loss history is needed to estimate the initial masses of very massive stars, especially for observations beyond a million years, as different initial masses converge to similar stellar masses.

\begin{deluxetable}{llll}
\tablenum{3}
\tablecaption{Luminosities and masses of the central WN5h stars}
\tablewidth{0pt}
\tablehead{
  \colhead{WB85$^{a}$} &
  \colhead{log\,$L$} &
  \colhead{Mass} & \colhead{Reference${^b}$} \\ \colhead{}&\colhead{($L_{\odot}$)} & \colhead{($M_{\odot}$)} & \colhead{} 
}
 \startdata
  R136\,a1 & 6.67$\pm$0.13 & 196$^{+34}_{-27}$ & This work \\
  & 6.79$\pm$0.1 & 251$^{+48}_{-31}$ & \cite{Best20} \\
  & 6.94$\pm0.09$ & 315$^{+60}_{-50}$ & \cite{Crow16} \\
  \hline
  R136\,a2 & 6.55$\pm$0.13 & 151$^{+27}_{-16}$ & This work\\
  & 6.75$\pm$0.1 & 211$^{+31}_{-32}$ & \cite{Best20} \\
  & 6.63$\pm0.09$ & 195$^{+35}_{-30}$ & \cite{Crow16}  \\
\hline  
  R136\,a3 & 6.56$\pm0.13$ & 155$^{+25}_{-18}$ & This work\\
    & 6.63$\pm$0.1 & 181$^{+29}_{-31}$ & \cite{Best20} \\
  & 6.58$\pm$0.09 & 180$\pm30$ & \cite{Crow16}  \\
\enddata
\tablecomments{(a) Based on the nomenclature from \cite{Baier} (b) All spectral parameters except luminosity are the same with \cite{Best20}. \cite{Crow16} adopted effective temperatures of 53$\pm$3\,kK for all sources.  }
\end{deluxetable}

\section{Conclusions}
In this paper, we have presented the sharpest optical images of the Large Magellanic Cloud cluster R136 taken using the speckle imager Zorro mounted on Gemini South Observatory. Our main conclusions are--

\begin{itemize}
\item Stellar census- we identify all known stars brighter than $V=16$\,mag within 2$\arcsec$ from R136a1 at angular resolutions between 30-40{\it mas} ($\sim$2000\,au at the adopted distance) in filters with central wavelengths similar to {\it BVRI}. The coordinates, and magnitudes of all sources detected are tabulated.

\item Cluster core- We resolve the central core surrounding the WN5h star R136a1, separating a1 from it's visual companion HSH17 (at an angular distance of 3000\,au). HSH17 is a mid-O spectral type star. The WN5h star R136a3 is also shown to have a faint visual companion in the optical $\lesssim$2000\,au away. These companions have minor probabilities of being chance alignments. 

\item The masses of the three central WN5h stars in R136 (a1, a2, a3) estimated from our photometry is between 200--150\,$M_{\odot}$-- significantly lower than the range of $\sim$320--180\,$M_{\odot}$ quoted in the literature. Since a1 is currently thought to be the most massive star known, our result indicates that the current empirical upper-mass limit of stellar initial masses is lower than previously claimed.

\end{itemize}

As concluding remarks, we caution that observations of this nature, to the authors knowledge have not been conducted in the literature, and push the boundary of what is considered possible using speckle photometry. For this reason, we express caution when interpreting our results, and use them only to suggest that currently, the validity of the known empirical upper-mass limit in this region previously found in the literature should only be considered as an estimate, and future much higher angular resolution images are essential to estimate the stellar luminosity (and mass) that can separate the flux contribution from various neighboring components.

\begin{acknowledgments}
V.M.K. thanks Asha Kalari for providing valuable feedback on the figures. We thank R{\'e}ne Rutten for facilitating the first test images, and Janice Lee for awarding directors time to our project. We are grateful to the Zorro team led by Steve Howell, and Gemini staff for taking these observations. V.M.K. thanks Drs. Chris Evans, Zainab Khorrami, and Norberto Castro for providing the nIR AO images, and MUSE AO images respectively. V.M.K. and R.S. are supported by the international Gemini Observatory, a program of NSF’s NOIRLab, which is managed by the Association of Universities for Research in Astronomy (AURA) under a cooperative agreement with the National Science Foundation, on behalf of the Gemini partnership of Argentina, Brazil, Canada, Chile, the Republic of Korea, and the United States of America. V.M.K. acknowledges funding from the Gemini-CONCIYT fellowship 32RF180005. M.R. wishes to acknowledge support from ANID(CHILE) through FONDECYT grant No. 1190684 and partial support from ANID project Basal AFB-170002. We thank the anonymous referee for a detailed report which helped improve the manuscript. 
\end{acknowledgments}

\vspace{5mm}
\facilities{Gemini}





\bibliography{sample631}{}
\bibliographystyle{aasjournal}



\end{document}